\documentclass[aps,superscriptaddress,showpacs,floatfix]{revtex4}
\usepackage{graphicx}
\usepackage{dcolumn}
\usepackage{bm}

\begin{document}

\title{Three-body force effect on neutrino emissivities of neutron stars within the framework of the Brueckner-Hartree-Fock approach}

\author{Peng Yin}
\affiliation{Institute of Modern Physics, Chinese Academy of
Sciences, Lanzhou 730000, China} \affiliation{University
of Chinese Academy of Sciences, Beijing, 100049, China}

\author{Wei Zuo \footnote{Corresponding author: zuowei@impcas.ac.cn}}
\affiliation{Institute of Modern Physics, Chinese Academy of
Sciences, Lanzhou 730000, China}\affiliation{Kavli Institute for Theoretical Physics, Chinese Academy of
Sciences, Beijing 100190, China}
\affiliation{State Key Laboratory of Theoretical Physics,
 Institute of Theoretical Physics, Chinese Academy of Sciences, Beijing 100190, China}

\begin{abstract}
The three-body force (TBF) effect on the neutrino emissivity
 in neutron star matter and the total neutrino emissivity of neutron stars have been investigated within
 the framework of the Brueckner-Hartree-Fock approach by adopting
the AV18 two-body interaction plus a microscopic TBF.
The neutrino emissivity from the direct Urca process turns out to be much larger than that from the modified Urca process.
Inclusion of the TBF reduces strongly the density thresholds of the direct Urca processes involving electrons and muons.
The TBF effect on the total neutrino emissivity of neutron stars is shown to be negligibly weak for neutron stars with small masses.
For neutron stars with large masses, the TBF effect becomes visible and inclusion of the TBF may enhance the total neutrino emissivity by about $50\%$ for neutron stars with a given mass of $M=1.6M_{\odot}$.

\end{abstract}

 \maketitle

\section{Introduction}

Neutron stars provide natural astrophysical laboratories for probing the properties of superdense matter. Modeling the cooling of neutron stars is one of the potential methods of investigating the structure of neutron stars~\cite{pethick:1992,page:1992,page:2004,yakovlev:2004}. Neutrino emissivity of a neutron star is of a great physical interest since it is one of the most important physical inputs to solve two equations: the thermal balance equation and the thermal transport equation~\cite{thorne:1977}, which are used to get the cooling curves of neutron stars~\cite{page:1992,page:2004,lattimer:1994}.

The standard scenario of the neutron star cooling is based on the modified Urca (MU) process. The MU process was first introduced in Ref.~\cite{chiu:1964} in the context of neutron star cooling. It is believed that the MU process can go through two channels:
\begin{eqnarray}\label{eq:MU}
n+n&\longrightarrow &p+n+l+\bar{\nu}_l,\ \ p+n+l\longrightarrow n+n+\nu_l;\\
n+p&\longrightarrow &p+p+l+\bar{\nu}_l,\ \ p+p+l\longrightarrow n+p+\nu_l,
\end{eqnarray}
which we define as the neutron and proton branches of the MU process, respectively. By the label $l$ we denote the leptons considered here, i.e., electrons and muons. The neutrino cooling rate in the neutron branches was calculated in Refs.~\cite{bahcall:1965,friman:1979,maxwell:1987}. The proton branch of the MU process was pointed out in Ref.~\cite{itoh:1972}. The energy loss rate of the proton branch was calculated in Ref.~\cite{maxwell:1987}, which shows that the proton branch is negligible compared with the neutron one. However, the authors of Ref.~\cite{yakovlev:1995} pointed out that the proton branch is as efficient as the neutron one. In the non-standard scenario of neutron star cooling, the most powerful neutrino reactions are expected to be the so-called direct Urca (DU) processes~\cite{pethick:1992,prakash:1992,prakash:1994}:
\begin{eqnarray}\label{eq:DU}
n\longrightarrow p+l+\bar{\nu}_l,\ \ p+l\longrightarrow n+\nu_l.
\end{eqnarray}
It has been pointed out in Ref.~\cite{lattimer:1991} that the $\beta$-stable matter with a proton-to-neutron ratio in excess of some critical value lying in the range of 11-15\% can cool by the DU process. Furthermore, the authors of Ref.~\cite{prakash:1992} have shown that matter with any proton-to-neutron ratio can rapidly cool by the DU process if $\Lambda$ hyperons are present.

The Brueckner-Hartree-Fock (BHF) approach is an approach based on the microscopic theory. It has been well developed and has also been widely used for
 predicting the properties of nuclear matter~\cite{jeukenne:1976,bombaci:1991,baldo:1999,zuo:1999}. Furthermore, the BHF approach has also been applied to investigate the structure of neutron stars~\cite{baldo:1997,zhou:2004,li:2008}, the neutrino emissivity inside neutron star cores~\cite{elgaroy:1996} as well as other properties of neutron stars~\cite{zuo:2004a,zuo:2004b,lombardo:2001,lombardo:2003}. Within the framework of the BHF approach, the TBF effect on the structure properties of neutron stars have been studied in Refs.\cite{zhou:2004,li:2008},
and it is shown that the TBF effect may increase remarkably the predicted maximum mass of neutron stars.
In the present paper, we investigate the neutrino emissivities of different processes in $\beta$-equilibrium neutron star matter and the total neutrino emissivity of neutron stars within the framework of the BHF method plus a microscopic three-body force (TBF). Particularly, we concentrate on the effect of the TBF on the neutrino emissivity
in the interior of neutron stars.

The present paper is organized as follows. In the next section, we
give a brief review of the adopted theoretical approaches including
the BHF theory and the TBF model. Furthermore, we briefly show formulas adopted here to calculate
the neutrino emissivties of different reactions in neutron stars.
In Sec. III, the calculated results will be reported and discussed. Finally, a summary is given in Sec. IV.

\section{Theoretical Approaches}
\subsection{The BHF approach}
The theoretical description of neutron star matter adopted in the present paper is the BHF
approach for asymmetric nuclear matter~\cite{zuo:1999}.
The BHF approach is based on
the reaction $G$-matrix, which satisfies the following isospin-dependent Bethe-Goldstone (BG) equation~\cite{day}:
\begin{eqnarray}\label{eq:BG}
G(n,\beta;\omega)= V_{NN}+
%\nonumber\\
 V_{NN}\sum\limits_{k_{1}k_{2}}\frac{|k_{1}k_{2}\rangle
Q(k_{1},k_{2})\langle
k_{1}k_{2}|}{\omega-\epsilon(k_{1})-\epsilon(k_{2})}G(n,\beta;\omega),
%\nonumber\\
\end{eqnarray}
where $k_i\equiv(\vec k_i,\sigma_i,\tau_i)$ denotes the momentum, and 
the $z$ components of spin and isospin of a nucleon, respectively.
$V_{NN}$ is the realistic $NN$ interaction and $\omega$ is the starting energy;
$\beta$ is isospin asymmetry of asymmetric nuclear matter and is defined as
$\beta=(n_{n}-n_p)/n$, where $n$, $n_n$ and $n_p$ denote
the total nucleon, neutron and proton
number densities, respectively. The Pauli operator $Q(k_{1},k_{2})$ prevents
two nucleons in intermediate sates from being scattered into their
respective Fermi seas.
$\epsilon(k)$ denotes the single-particle (s.p.) energy  which is given by:
\begin{eqnarray}\label{eq:SPE}
 \epsilon(k)=\hbar^{2}k^{2}/(2m)+U(k),
\end{eqnarray}
where $U(k)$ is the auxiliary s.p. potential which controls the convergent rate of the
hole-line expansion~\cite{day}.
The continuous choice is adopted for the auxiliary potential in the present calculation,
which will result in a much faster convergence of
the hole-line expansion up to high densities than the gap choice~\cite{song:1998}.
 The s.p. potential under the continuous choice describes physically the nuclear mean
 field felt by a nucleon in nuclear medium at the lowest BHF level
~\cite{lejeune:1978} and is
calculated via the relation:
\begin{eqnarray}\label{eq:UBHF}
U(k)=\rm{Re}\sum\limits_{k'\leq k_{F}}\langle
kk'|G[n,\beta;\epsilon(k)+\epsilon(k')]|kk'\rangle_{A} \ ,
\end{eqnarray}
where the subscript $A$ denotes anti-symmetrization of the matrix
elements.

We adopt the Argonne $V_{18}$ (AV18) two-body interaction~\cite{wiringa:1995} plus a
microscopic TBF~\cite{zuo:2002a} for the realistic $NN$ interaction $V_{NN}$.
The TBF model adopted here is constructed by using the meson-exchange current
approach~\cite{grange:1989} and the most important mesons, i.e., $\pi$, $\rho$,
 $ \sigma $ and $\omega$ have been considered.
The parameters of the TBF model, i.e., the coupling constants
and the form factors, have been self-consistently determined to reproduce the AV18
two-body force using the one-boson-exchange potential
model and their values can be found in Ref.~\cite{zuo:2002a}.
 In the present calculation, the TBF has been reduced to an
equivalently effective two-body interaction according to the
standard scheme as described in Ref.~\cite{grange:1989} and
 the corresponding effective two-body force $V_3^{\rm eff}$ in $r$-space is given as follows:
\begin{eqnarray}\label{eq:tbf}
 \langle \vec r_1^{\ \prime} \vec r_2^{\ \prime}| V_3^{\rm eff} |
\vec r_1 \vec r_2 \rangle = \displaystyle
 \frac{1}{4} {\rm Tr} \sum_{m} \int {\rm d}
{\vec r_3} {\rm d} {\vec r_3^{\ \prime}}\phi^*_m(\vec r_3^{\
\prime}) [1-\eta(r_{13}')] [1-\eta(r_{23}')] \nonumber \\
\times W_3(\vec r_1^{\ \prime}\vec r_2^{\ \prime} \vec r_3^{\
\prime}|\vec r_1 \vec r_2 \vec r_3)
 \phi_m(\vec r_3)
[1-\eta(r_{13})]
 [1-\eta(r_{23})].
\end{eqnarray}

In the BHF approximation, the energy per nucleon of asymmetric nuclear matter is given by
\begin{eqnarray}\label{eq:BHFE}
E_{A}(n,\beta)=\frac{3}{5}\frac{\hbar^2}{2m} [(\frac{1-\beta}{2})^{5/3}+(\frac{1+\beta}{2})^{5/3}] (3\pi^2 n)^{2/3}
+\frac{1}{2n} \rm{Re} \sum_{\tau,\tau'}
\sum_{k\leq k^{\tau}_F,k'\leq k^{\tau'}_F}\langle kk'|G[n,\beta;\epsilon(k)+\epsilon(k')]|kk'\rangle_A.
\end{eqnarray}
where the first term is the contribution of the kinetic part; the second term is the potential part.
It has been demonstrated by microscopic calculations~\cite{bombaci:1991,zuo:1999,zuo:2002b,engvik:1996,baldo:1998,dalen:2005} that
the energy per nucleon of asymmetric nuclear matter fulfills satisfactorily a quadratic dependence on asymmetry $\beta$ in the whole asymmetry
range of $0\le\beta\le1$, i.e., $E_{A}(n,\beta)=E_{A}(n,\beta=0)+\beta^2E_{sym}(n)$,
where the symmetry energy $E_{sym}(n)$ describes the isospin dependence of the equation of state (EOS) of asymmetric nuclear matter,
and it plays a decisive role in determining the proton fraction $x\equiv n_p/n =(1-\beta)/2$ inside $\beta$-stable ($npe\mu$) neutron star matter \cite{lattimer:1991,bombaci:1991}. As soon as the $G$ matrix is obtained by solving the BG equation self-consistently with the s.p. potential, one can
calculate the EOS of asymmetric nuclear matter and the symmetry energy. With the aid of the $\beta$-equilibrium and charge
neutrality conditions, the EOS of $\beta$-stable ($npe\mu$) neutron star matter can be readily obtained. By solving the Tolman-Oppenheimer-Volkov (TOV) equation, one may determine finally the properties of neutron stars, including the mass-radius relation, the neutron and proton distributions inside a neutron star with a given mass, etc.

\subsection{Neutrino emissivities of the DU process and the MU process}
The derivation of the neutrino emissivity of the DU process under the condition of $\beta$ equilibrium is based on the $\beta$ decay theory and more details can be found in Ref.~\cite{yakovlev:2001}. The neutrino emissivity of the DU process in the $npe$ neutron star matter is calculated as follows~\cite{lattimer:1991}:
\begin{eqnarray}\label{eq:DQ}
Q^{(D)}=4.00\times10^{27}\left(\frac{n_e}{n_0}\right)^{1/3}\frac{m_n^*m_p^*}{m_n^2}T_9^6\Theta_{npe} \ \rm{erg}\  \rm{cm}^{-3}\rm{s}^{-1},
\end{eqnarray}
where $m_n^*$ and $m_p^*$ denote the neutron and proton effective masses, respectively, and they are calculated by the BHF approach in the present paper. $T_9$ is the temperature in units of $10^9$K and $\Theta_{npe}$ is the step function: $\Theta_{npe}=1$ if the Fermi momenta $p_{F_n},p_{F_p},p_{F_e}$ satisfy the triangle condition, i.e., $p_{F_p}+p_{F_e}>p_{F_n}$, and $\Theta_{npe}=0$ otherwise.

If muons are present, then the DU process involving muons may become possible along with the process involving electrons. Its emissivity is given by the same Eq.~(\ref{eq:DQ}) except that $\Theta_{npe}$ must be replaced by $\Theta_{np\mu}$. Thus the neutrino emissivity is increased by a factor of 2 accordingly if the DU process involving muons is allowed~\cite{lattimer:1991}.

The practical expression for calculating the emissivity of MU process has been introduced in Ref.~\cite{friman:1979}. The emissivity in the neutron branch is given as follows:
\begin{eqnarray}\label{eq:MNQ}
Q^{(Mn)}=8.55\times10^{21}\left(\frac{m_n^*}{m_n}\right)^{3}\left(\frac{m_p^*}{m_p}\right)
\left(\frac{n_e}{n_0}\right)^{1/3}T_9^8\alpha_n\beta_n \ \rm{erg}\  \rm{cm}^{-3}\rm{s}^{-1},
\end{eqnarray}
where the factor $\alpha_n$ describes the momentum transfer dependence of the squared reaction matrix element
of the neutron branch under the Born approximation, and $\beta_n$ includes the non-Born
corrections and the corrections due to the $NN$ interaction effects which are not described by the one-pion exchange
 and/or the Landau theory as mentioned in Ref.~\cite{yakovlev:1995}. In the present paper, we adopt the same values of $\alpha_n$ and $\beta_n$ as in Refs.~\cite{friman:1979,yakovlev:1995}, i.e., $\alpha_n=1.13$
 and $\beta_n=0.68$.

 The expression for the neutrino emissivity in the proton branch is given by~\cite{yakovlev:1995}:
 \begin{eqnarray}\label{eq:MPQ}
Q^{(Mp)}=8.55\times10^{21}\left(\frac{m_p^*}{m_p}\right)^{3}\left(\frac{m_n^*}{m_n}\right)
\left(\frac{n_e}{n_0}\right)^{1/3}T_9^8\alpha_p\beta_p\left(1-\frac{p_{F_e}}{4p_{F_p}}\right)\Theta_{Mp}\ \rm{erg}\  \rm{cm}^{-3} \rm{s}^{-1},
\end{eqnarray}
where we set $\alpha_p=\alpha_n$ and $\beta_p=\beta_n$. The above expression is similar to Eq.~(\ref{eq:MNQ}) and the main difference between the proton branch and the neutron branch is the threshold character: The proton branch is allowed if $p_{F_n}<3p_{F_p}+p_{F_e}$, i.e., $\Theta_{Mp}=1$ only under the condition of $p_{F_n}<3p_{F_p}+p_{F_e}$ and $\Theta_{Mp}=0$ otherwise. By comparing the expression for $Q^{(Mn)}$ and $Q^{(Mp)}$, one can easily obtain $Q^{(Mp)}/Q^{(Mn)}=(m_p^*/m_n^*)^2[1-p_{F_e}/(4p_{F_p})]$.

If muons are present in the dense neutron star matter, the MU process involving muons becomes possible. Accordingly, several modifications should be included in Eqs.~(\ref{eq:MNQ}) and (\ref{eq:MPQ}) as discussed in Ref.~\cite{yakovlev:2001}. First, $\Theta_{Mp}$ and $p_{F_e}$ should be replaced by $\Theta_{Mp\mu}$ (muon threshold function) and $p_{F_\mu}$, respectively. Second, an additional factor $v_{F_\mu}/c=(n_\mu/n_e)^{1/3}$ should be included in both expressions, where $v_{F_\mu}$ is the Fermi velocity of muons. In addition, the muon neutron branch can not be switched on below the threshold density $\rho_\mu$ at which muons appear and the threshold density of the muon proton branch, accounted by the step function $\Theta_{Mp\mu}$, is also naturally restricted by $n>\rho_\mu$.

According to the discussion above, the neutrino emissivity jumps directly from the value of the MU process to that of the DU process. Thus, the DU process appears in a step-like manner. However, there should be a thermal broadening of the DU threshold for the practical situation as discussed in Ref.~\cite{yakovlev:2001}. In the present investigation, the thermal broadening is omitted since it is expected to be weak for many applications.

\section{Results and Discussion}

\begin{figure}[tbh]
\begin{center}
\includegraphics[width=0.6\textwidth]{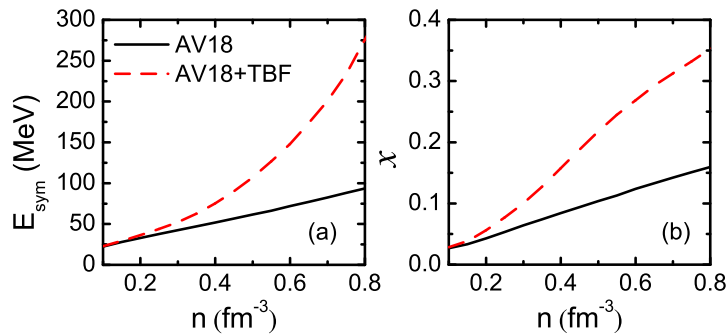}
\end{center}
\caption{(Color online) TBF effect on the symmetry energy (left panel) and the proton fraction (right panel).}\label{fig1}
\end{figure}

In Fig.\ref{fig1} we display the TBF effect on the calculated symmetry energy (left panel) and the
proton fraction (right panel) in $\beta$-stable neutron star matter as functions of nucleon number density.
In the figure, the solid lines are obtained by adopting
purely the AV18 two-body force alone; the dashed ones correspond to the results obtained by including the TBF.
We notice from Fig.~\ref{fig1} that including TBF increases significantly the symmetry energy at high densities and leads to a strong stiffening of the density dependence of symmetry energy at supra-saturation density as also shown in Refs.~\cite{zuo:2002b,li:2006}.
As a consequence, the proton fraction in the interior of neutron stars predicted by including the TBF is much larger
than that of excluding the TBF~\cite{zhou:2004}. One may also notice that in both the calculations with and without TBF,
the proton fraction $x$ increases monotonically as a function of nucleon number density and
can exceed the threshold value, i.e., $x^{DU}=(11\rm{-}15)\%$, above which the DU process
is allowed as mentioned before, whereas, inclusion of the TBF shifts the threshold nucleon densities from $n_e^{DU}=0.67 \, \rm{fm}^{-3}$ and
 $n_\mu^{DU}=0.76 \, \rm{fm}^{-3}$ to much lower values of $n_e^{DU} = 0.38 \, \rm{fm}^{-3}$ and $n_\mu^{DU} = 0.42 \, \rm{fm}^{-3}$
 for the electron and muon DU processes, respectively.

\begin{figure}[tbh]
\begin{center}
\includegraphics[width=0.5\textwidth]{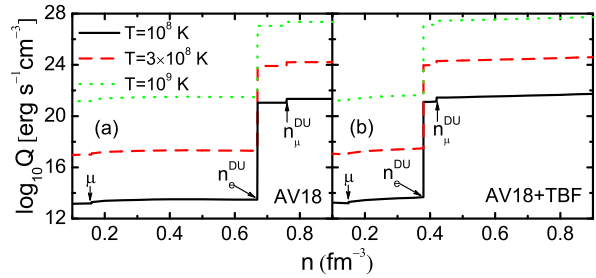}
\end{center}
\caption{(Color online) Density dependence of the neutrino emissivity in $npe\mu$ matter at three different temperatures,i.e.,$T=10^8, 3\times10^8$, and $10^9$ K. The left panel is calculated by adopting purely the AV18 two-body force and the right one is obtained by including the TBF. Arrows mark the densities where muons appear ($\mu$), and where the DU processes involving electrons ($n_e^{DU}$) and muons ($n_{\mu}^{DU}$) occur.}\label{fig2}
\end{figure}

In Fig.~\ref{fig2}, we report the dependence of the neutrino emissivity on the nucleon number density in neutron star matter.
In the figure, the TBF effect is also displayed. The left panel is obtained by adopting purely
the AV18 two-body interaction and the TBF is not included. Here we simply take the MU process and the DU process into account.
It can be seen from the left panel of Fig.\ref{fig2} that muons appear above the density of $n=0.16 \, \rm{fm}^{-3}$.
The DU processes involving electrons and muons are switched on at the threshold densities of $n_e^{DU}=0.67 \, \rm{fm}^{-3}$ and $n_\mu^{DU}=0.76 \, \rm{fm}^{-3}$, respectively. Therefore, the DU process are expected to occur in neutron stars with a central density higher than $0.67 \, \rm{fm}^{-3}$. At $n < 0.16 \, \rm{fm}^{-3}$, the neutrino emissivity is completely determined by the electron MU process. At densities of $n>0.16 \, \rm{fm}^{-3}$, the presence of muons enhances slightly the neutrino emissivity. Above $n>0.67 \, \rm{fm}^{-3}$, the onset of the DU process turns out to amplify the neutrino emissivity by 6-8 orders of magnitude, in agreement with the previous investigations~\cite{yakovlev:2001}. Therefore, one can expect that the DU process plays the predominant role in determining the neutron star cooling over the MU process~\cite{yakovlev:2001,lattimer:1991,lattimer:1994,lattimer:2004,gnedin:2001}.

The right panel of Fig.~\ref{fig2} is obtained by adopting the AV18 plus the TBF.
By comparing the results in the left and right panels, one may notice that the most significant effect of the TBF is to reduce the threshold densities for the onset of the electron and muon DU processes to much lower values, i.e., from $n_e^{DU}=0.67 \, \rm{fm}^{-3}$ and
 $n_\mu^{DU}=0.76 \, \rm{fm}^{-3}$ down to $n_e^{DU} = 0.38 \, \rm{fm}^{-3}$ and $n_\mu^{DU} = 0.42 \, \rm{fm}^{-3}$, respectively.  As a result, the main difference between the neutrino emissivity predicted by including the TBF and that obtained by excluding the TBF lies in the density region of $0.38 \, \rm{fm}^{-3}<n<0.67 \, \rm{fm}^{-3}$ due to the different density thresholds of the DU processes in these two cases. It is clear from Fig.\ref{fig2} that, in the density range from $0.38$ fm$^{-3}$ to $0.67$ fm$^{-3}$, inclusion of the TBF leads to a strong enhancement of the calculated neutrino emissivity by 6-8 orders of magnitude in neutron star matter at a given temperature.
 As it has been pointed out in Ref.~\cite{lattimer:1991}, the more rapidly the symmetry energy increases with density, the lower the density at which the DU process begins to operate. Therefore, the above-resulting TBF effect is expected since inclusion of the TBF stiffens the density dependence of symmetry
 energy and increases strongly the proton fraction in neutron star matter at high densities as displayed in Fig.~\ref{fig1}.
 In other density ranges ($n<0.38$ fm$^{-3}$ and $n>0.67$ fm$^{-3}$), the neutrino emissivity is enhanced slightly by inclusion of the TBF compared with the case of excluding the TBF.
\begin{table}\begin{center}

\caption{\label{tab:table1} Properties of neutron stars with three different masses ($1.2M_\odot$,$1.4M_\odot$,$1.6M_\odot$) obtained by adopting the AV18 two body force including and excluding the TBF: $n_c$ denotes the central density, $R$ is the radius of neutron star, and $R^{DU}$ is the radius range in which the DU processes are allowed.}
\begin{tabular}{|c|c|c|c|c|c|c|}\hline
\multicolumn{1}{|c|}{ } & \multicolumn{2}{c|}{$n_c (\rm{fm}^{-3})$} & \multicolumn{2}{c|}{R (\rm{km})}& \multicolumn{2}{c|}{$R^{DU} (\rm{km})$}\\
\cline{2-7}
&AV18 &AV18+TBF
&AV18 &AV18+TBF
&AV18 &AV18+TBF\\ \hline
$1.2M_\odot$&0.75&0.40&10.11&12.32&3.60&3.47\\ \hline
$1.4M_\odot$&0.92&0.44&9.86&12.31&5.40&5.69\\ \hline
$1.6M_\odot$&1.19&0.50&9.37&12.27&6.40&7.09\\ \hline
\end{tabular}
\end{center}
\end{table}

In Fig.~\ref{fig2}, we have displayed the density dependence of the neutrino emissivity
obtained by adopting the AV18 two-body force with and without the TBF. Thus one can roughly
estimate the total neutrino emissivity of a neutron star with a given mass ($Q_t$) by the following
integral once the density distribution in the neutron star is obtained:
 \begin{eqnarray}\label{eq:ENC}
 Q_t=\int_{0}^{R}4\pi r^2Q(r)dr,
\end{eqnarray}
where $R$ denotes the radius of a neutron star with a given mass and $Q(r)$ is the neutrino emissivity at the
location $r$ inside the neutron star (i.e., the radial distance from the center of the neutron star).
In order to calculate the density distribution of a neutron star, one has to solve the well-known TOV
equations~\cite{shapiro:1983}. In Table~\ref{tab:table1}, we display several properties of neutron stars
with three different masses obtained by solving the TOV equations, where the physics inputs are calculated
within the BHF approach by adopting the AV18 pure two-body force and the AV18 two-body force plus the TBF.
We can see from Table~\ref{tab:table1} that
inclusion of the TBF reduces greatly the central density of a neutron star with a fixed mass as compared with the case
of excluding the TBF. Besides, the radii of neutron stars with certain masses obtained by adopting AV18+TBF turn out to be
much larger than the corresponding ones obtained by adopting purely the AV18 two-body force.
The above results are consistent with the results in Ref.~\cite{baldo:1997} within the BHF approach by adopting
the Argonne AV14 and the Paris two-body force, implemented by the Urbana model for the TBF. Therefore one may
expect that the density inside a neutron star obtained by adopting the pure AV18 two-body force, decreases more rapidly
than that calculated by including the TBF going from the center to the crust. In Table~\ref{tab:table1}, we also
report the radius range $R^{DU}$ of neutron star where the DU processes are allowed, i.e., the DU processes
can occur from the center of neutron star up to the radius of $R^{DU}$. One may notice that the TBF effect on the range $R^{DU}$
is quite weak for neutron stars with
relatively small masses. For the neutron star with a larger mass of $M=1.6M_{\odot}$, inclusion of the TBF tends to increase $R^{DU}$ by about $10\%$
from 6.40 to 7.09 km.

\begin{figure}[tbh]
\begin{center}
\includegraphics[width=0.5\textwidth]{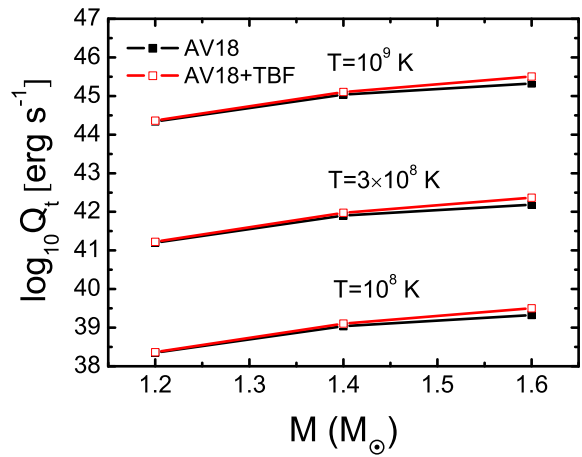}
\end{center}
\caption{(Color online) Total neutrino emissivity of neutron stars for three different masses ($1.2M_\odot,1.4M_\odot,1.6M_\odot$) at three typical temperatures ($10^8 \rm{K},3\times10^8 \rm{K},10^9 \rm{K}$). The filled squares are calculated by adopting purely the AV18 two-body force and the open ones are obtained by including the TBF.} \label{fig3}
\end{figure}

In Fig.~\ref{fig3}, we show the total neutrino emissivity of neutron stars with certain masses
at three different temperatures, calculated by Eq.~(\ref{eq:ENC}).
It is noticed from Fig.~\ref{fig3} that the total neutrino emissivity from a neutron star
increases greatly as a function of the neutron star mass. This is readily understood from the mass dependence
of the radius range $R^{DU}$ for the DU processes. It is easily seen from Fig.~\ref{fig2} that the total neutrino emissivity
in Eq.~(\ref{eq:ENC}) is contributed mainly from the DU processes, i.e., the radius range $R^{DU}$ plays the dominant role
in the integral of Eq.~(\ref{eq:ENC}). We can notice from Table~\ref{tab:table1} that the radius range $R^{DU}$ increases
greatly as the mass of neutron star increases for the case of either including or excluding the TBF. Thus we
can explain the mass dependence of the total neutrino emissivity.
In Fig.~\ref{fig3}, we also investigate the TBF effect on the total neutrino emissivity of neutron stars.
By comparing the filled squares and the corresponding open ones in Fig.~\ref{fig3},
one may notice that the TBF effect is negligibly small for neutron stars with small masses around $1.2M_\odot$.
For neutron stars with larger masses, the TBF effect becomes visible.
It is seen that the TBF effect is to enhance the total neutrino emissivity from neutron stars.
It is worth noticing that although the TBF leads to a remarkable enhancement of the proton fraction and a strong reduction of the threshold density
for the DU processes in $\beta$-stable neutron star matter,
its effect on the total neutrino emissivity from neutron stars is not very large as expected.
For example, inclusion of the TBF enhances the total neutrino emissivity by only about $50\%$ 
for a neutron star with $M=1.6M_{\odot}$.
The above results concerning the TBF effect on the total neutrino emissivity can be easily understood according to the TBF effect on the range $R^{DU}$ for the DU processes in neutron stars.  As has been discussed in connection with Table~\ref{tab:table1}, the TBF affects the predicted $R^{DU}$ only slightly for neutron stars with relatively low masses of less than 1.4$M_{\odot}$. Its effects increase as the neutron star mass increases and it enhances the $R^{DU}$ by about $10\%$ as the neutron star mass increases to 1.6$M_{\odot}$.

\section{Summary}

In summary, we have investigated the TBF effect on the neutrino emissivity in the interior of neutron stars within the framework of the BHF approach by adopting the AV18 two-body interaction supplemented with a microscopic TBF.
 It is shown that the neutrino emissivity from the DU process is much larger than that from the MU process, in agreement with the previous predictions~\cite{yakovlev:2001,lattimer:1991,lattimer:1994,lattimer:2004,gnedin:2001}.
The TBF is expected to increase the proton fraction in the $\beta$-stable neutron star matter~\cite{zhou:2004,li:2008}, and consequently to reduce remarkably the density thresholds of the DU processes involving electrons and muons from $n_e^{DU}=0.67 \, \rm{fm}^{-3}$ and
 $n_\mu^{DU}=0.76 \, \rm{fm}^{-3}$ to $n_e^{DU} = 0.38 \, \rm{fm}^{-3}$ and $n_\mu^{DU} = 0.42 \, \rm{fm}^{-3}$, respectively. Furthermore, it is shown that inclusion of the TBF enhances strongly the neutrino emissivity in the density region of $0.38 \, \rm{fm}^{-3}<n<0.67 \, \rm{fm}^{-3}$ in $\beta$-stable neutron star matter. In the density region out of $0.38 \, \rm{fm}^{-3}<n<0.67 \, \rm{fm}^{-3}$, the TBF effect on the neutrino emissivity turns out to be quite small.
In addition, we have investigated the TBF effect on the total neutrino emissivity from neutron stars.
 The radius range $R^{DU}$ , in which the DU processes are possible, is calculated. It is found that
 the radius range $R^{DU}$ plays the dominant role in determining the total neutrino emissivity from neutron stars and it
 increases considerably as the neutron star mass increases. As a consequence, the total neutrino emissivity from a neutron star increases as its mass increases. For neutron stars with small masses, the TBF effect on the total neutrino emissivity is proved to be negligibly small. The TBF effect on the total neutrino emissivity becomes visible for neutron stars with relatively large masses. However, it is not as strong as expected and it leads an enhancement of the total neutrino emissivity of a neutron star with a given mass of 1.6$M_{\odot}$ by only about $50\%$.

\section*{Acknowledgments}

{The work is supported by the 973 Program of China under No. 2013CB834405,
the National Natural Science Foundation of China
(No. 11175219), and the Knowledge
Innovation Project(No. KJCX2-EW-N01) of the Chinese Academy of Sciences.}

%\end{CJK*}
\end{document}